\documentclass{llncs}
\usepackage{llncsdoc}
\usepackage{aliascnt}
\usepackage{remreset}
\usepackage{graphicx}
\graphicspath{{images/}}
\usepackage{wrapfig}
\usepackage[hyphens]{url}
\usepackage{hyperref}
\hypersetup{
    colorlinks=true,
    linkcolor=blue,
    urlcolor=blue,
    citecolor=blue,
}
\usepackage[hypcap]{caption}
\usepackage[T1]{fontenc}
\begin{document}

\title{ArchiveWeb: Collaboratively Extending and Exploring Web Archive
  Collections} 
\author{Zeon Trevor Fernando \and Ivana Marenzi \and Wolfgang Nejdl
  \and Rishita Kalyani}  
\institute{L3S Research Center, Hannover, Germany\\
\email{\{fernando,marenzi,nejdl,kalyani\}@L3S.de}
}

\maketitle

\begin{abstract}
  Curated web archive collections contain focused digital contents
  which are collected by archiving organizations to provide a
  representative sample covering specific topics and events to
  preserve them for future exploration and analysis. In this paper, we
  discuss how to best support collaborative construction and
  exploration of these collections through the ArchiveWeb
  system. ArchiveWeb has been developed using an iterative
  evaluation-driven design-based research approach, with considerable
  user feedback at all stages. This paper describes the
  functionalities of our current prototype for searching,
  constructing, exploring and discussing web archive collections, as
  well as feedback on this prototype from seven archiving
  organizations, and our plans for improving the next release of the
  system.  
\keywords{Working with Web Archives, Collaborative Search and Exploration}
\end{abstract}

\section{Introduction}\label{sec:Introduction}

The web reflects a considerable part of our society and is becoming an
important corpus for studying human society by researchers in the
humanities, social sciences, and computer sciences alike. Web archives
collect, preserve, and provide ongoing access to ephemeral web pages
and hence encode important traces of human thought, activity, and
history. Curated web archive collections contain focused digital
content from specific organizations, related to specific topics or
covering specific events, which are collected to provide
representative samples and preserve them for future exploration and
analysis. However, there have been only a few concerted efforts to
provide tools and platforms for exploring and working with such web
archives.

Within the ALEXANDRIA
project\footnote{\url{http://alexandria-project.eu/}} we aim to
develop models, tools and techniques not only to archive and index
relevant parts of the web, but also to retrieve and explore this
information in a meaningful way. This paper focuses on the ArchiveWeb
platform which supports searching, collecting, exploring and
discussing web archive collections such as the ones provided through the web
archiving service of the Internet Archive,
Archive-It\footnote{\url{http://archive-it.org/}}.

Archive-It provides a subscription web archiving service that helps
organizations to harvest, build, and preserve collections of digital
content. Currently the Archive-It system is mainly used by librarians
and curators in order to build their collections. Less support is
given to users and domain experts who actually want to work with
the collections, or to the general public. 
ArchiveWeb\footnote{\url{http://archiveweb.l3s.uni-hannover.de/aw/index.jsf}} aims to provide facilities to collaboratively explore
and work with web collections in an interactive and user friendly
way, both for research and for learning.

In summer 2015 we carried out a preliminary analysis of user
requirements during which we interviewed representatives of the
Internet Archive as well as experts working in seven archiving
organizations and libraries active in web archiving. In most cases the
motivation for starting the archiving of digital collections had been
to preserve institutional or government websites, or collecting and
preserving research outputs (for example faculty members hosting their
funded projects on external websites including assets related to the
project such as documents and research datasets). Several institutions
also curate special collections received as donations of materials
from individuals or organizations. Traditionally these materials were
paper manuscripts, but they increasingly include digital and
multimedia materials as well as social media and web presence.

All respondents confirmed that the current efforts at their
institutions are mainly carried out by individual curators or subject
specialists who are responsible for individual collections and
archiving requests. In some cases there is a collection development
executive (CDE) group, as part of the library organization, that
coordinates collection building and development.  Collaboration is an
important aspect in order to avoid duplicating resources and efforts,
one of the challenges being ``figuring out how to not duplicate
efforts that are going on elsewhere, but also how to still build
things that are useful''. Some curators are trying to find ways of
collaborating both in terms of collecting and displaying information,
sharing metadata and schema creation, user folksonomies and tagging,
to make the collections come together and be more alive. Easy annotation
of resources in collections is a feature that many experts would be
happy to explore.

On the basis of this preliminary study, and building on previous work
we did for collaborative learning environments ~\cite{Abel09,Marenzi2010,marenzi12,tlt12,oro42050} we
designed and built the ArchiveWeb system to support collaborative
creation and enrichment of web archive collections, with a focus on
user interface and searching/sharing functionalities. We ingested 200
web archive collections from Archive-It, and asked our archiving
partners for their input. Evaluation was done based on a task-based
evaluation design, and analyzing both quantitative interaction data
(query log), as well as qualitative feedback in the form of
interviews.

In the following section, we provide a short overview of related work,
and in Section \ref{sec:Web Archive Collections} we describe some web
archive collections we are working with. Section \ref{sec:System}
includes both a description of the main functionalities of our system
and of our evaluation design, a detailed discussion of the evaluation
results, as well as directions for future work.

\section{Related Work}\label{sec:Related Work}

Tools for supporting search and exploration in web archives are still
limited~\cite{dougherty09}. Most desired search functionalities in web
archives are fulltext search with good ranking, followed by URL
search~\cite{archiving07}. A recent survey showed that 89\% of web
archives provide URL search access and 79\% give metadata search
functionalities~\cite{gomes11}. Some existing projects that provide
limited support for web archive research are discussed below.

The Wayback Machine\footnote{\url{http://archive.org/web/}} is a web archive
access tool supported by the Internet Archive. It provides the ability
to retrieve and access web pages stored in a web archive through URL
search. The results for each URL are displayed in a calendar view
which displays the number of times the URL was crawled by the Wayback
Machine. Archive-It and ArchiveTheNet\footnote{\url{http://archivethe.net/}} are web
archive services provided by the Internet Archive and the Internet
Memory Foundation. These services enable focused archiving of web
contents by organizations, such as universities or libraries, that
otherwise could not manage their own archives.  
The Memento Project\footnote{\url{https://tools.ietf.org/html/rfc7089}} enables
the discovery of archived content from across multiple web archives
via URL search.

A few researchers have worked on providing new interfaces and
visualizations for searching, exploring and discovering insights from
web archives. Odijk et. al.~\cite{odijk15} present an exploratory
search interface to improve accessibility of digital archived
collections for humanities scholars, in order to highlight different
perspectives across heterogeneous historical collections. The
motivation for this work derives from the huge amount of digital
material that has become available to study our recent history,
including books, newspapers and web pages, all of which provide
different perspectives on people, places and events over time. In
their paper, the authors connect heterogeneous digital collections
through the temporal references found in the documents as well as
their textual content, in order to support scholars to detect,
visualize and explore materials from different perspectives. Padia
et. al.~\cite{padia12} provide an overview of a web archive
collection by highlighting the collection's underlying characteristics
using different visualizations of image plots, wordle, bubble charts
and timelines. Lin et. al.~\cite{lin14} present an interactive visualization based on topic models for exploring archived content,
so that users can get an overview of the collection content. The visualization displays a person-by-topic matrix that shows the association between U.S. senators websites and the derived topics. It also provides drill-down capabilities for users to examine the pages in which a topic is prevalent.

All of the above tools and interfaces help support the
exploration and search of web archives for individual users and
researchers. In addition, ArchiveWeb aims at
supporting the collaborative exploration of web archives. 
Previous
research on helping users keep track of their resources include tools
that provide better search and organizational facilities based on
metadata/time~\cite{dumais03} or tagging~\cite{cutrell06}. Our system provides similar organizational functionalities refined through several
learning communities and previous work, LearnWeb~\cite{tlt12}, thus gaining advantage from several years
of development and user feedback in that context. ArchiveWeb builds on the LearnWeb platform which already supports
collaborative sensemaking~\cite{Evans08,Russell93} by allowing users to share and
collaboratively work on resources retrieved from various web sources~\cite{Amershi08,Held2009,Morris2008,Morris2007}. 

\section{Web Archive Collections}\label{sec:Web Archive Collections}

Archive-It is a subscription web archiving service from the Internet
Archive that helps organizations to harvest, build, and preserve
collections of digital content. It was first deployed in 2006 and is
widely used as a service to collect, catalog, and manage
collections of archived web content. Fulltext search is also
available, even though effective ranking is still an open issue.  All
content is hosted and stored at the Internet Archive data centers.

Currently about 200 collections from Archive-It have been integrated
into the ArchiveWeb system with full metadata indexing, and can be
explored through a visual interface. Main topics covered by the
Archive-It collections available through ArchiveWeb are: Human Rights,
Contemporary art, Global Events as well as various Web resources  related to society, history, culture, science, statistics, and
governments of various countries.

\subsection{Human Rights}

\begingroup

\setlength{\intextsep}{5pt}
\begin{wrapfigure}{l}{0.25\textwidth}
\includegraphics[width=0.9\linewidth]{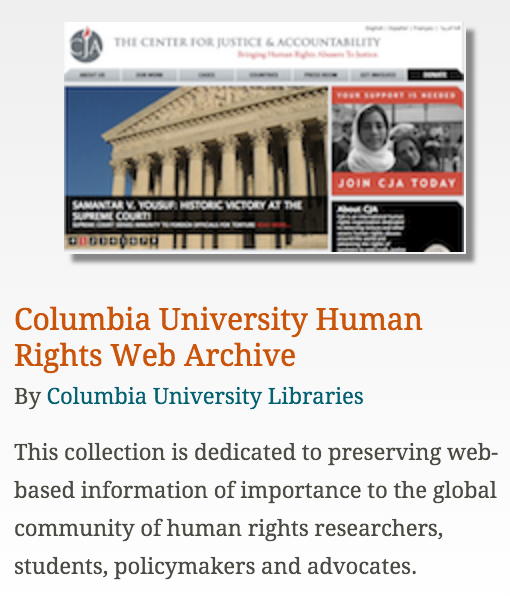} 
\label{hrwa}
\end{wrapfigure}

The collection \textit{Human Rights Web Archive} (HRWA) by
\textit{Columbia University Libraries} is made up of searchable
archived copies of websites related to human rights created by various
non-governmental organizations, national human rights institutions,
tribunals and individuals. The collection was started in 2008 and is still
being continued, adding new websites on a regular
basis. 

Identification of websites for archiving is done by subject
specialists with expertise in human rights and different regions of
the world at Columbia University Libraries. Public nominations are
provided by human rights researchers, advocates, organizations and
individuals who are involved in the creation of human rights related
websites. Priority is given to websites hosted in countries that do
not have any systematic web archiving initiatives in place. Websites
of intergovernmental organizations such as the United Nations are not
included in the collection.

Archive-It services are used to maintain the
collection, and the Internet Archive and Columbia University Libraries
store copies of the resulting data. The collection includes over \textit{711
  websites} with more than 50 million searchable documents and over
\textit{115 million archived documents} with an archived data size of
more than \textit{5TB}.

\endgroup

\begin{figure}[hbp]
\includegraphics{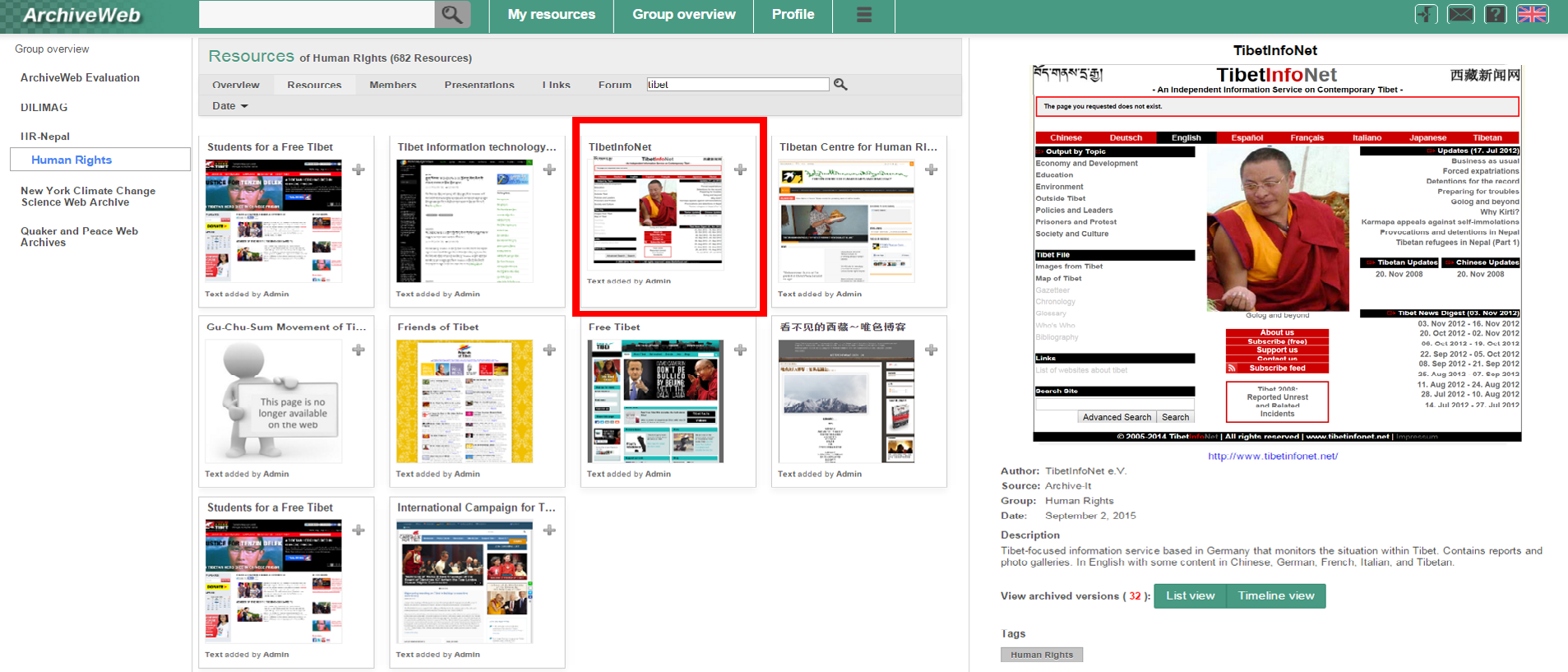}
\caption{Human Rights Example Resource}
\label{hrwa_ex}
\end{figure}

The HRWA collection provides a good balance between websites of large
and well known human organizations based in North America and Europe,
and websites of smaller organizations from other regions. It includes
websites from organizations such as Human Rights Watch, Amnesty
International, Transparency International, International Crisis Group,
and websites from regions with less web archiving activity and more
political/social unrest. It also covers websites and organizations
which are at greater risk of disappearing. One example of a website
that no longer has a live version but can still be accessed by
researchers via the Web archive collection is TibetInfoNet\footnote{http://www.tibetinfonet.net/}, which
monitors the situation in Tibet. This
page has been captured between May 2008 and July 2015, the ArchiveWeb
screenshot shows the human rights group restricted to ``Tibet''
related resources, with the TibetInfoNet resource on the right side
(Figure \ref{hrwa_ex}). 

The URL provided in ArchiveWeb below the screenshot on the right side
points to the original web site, which is no longer available on the
live web, but can be retrieved in its previous version through the
Wayback Machine (as described in section \ref{sec:enriching}).

\subsection{Occupy Movement 2011/12}

\begingroup

\setlength{\intextsep}{5pt}
\begin{wrapfigure}{l}{0.25\textwidth}
\includegraphics[width=0.9\linewidth]{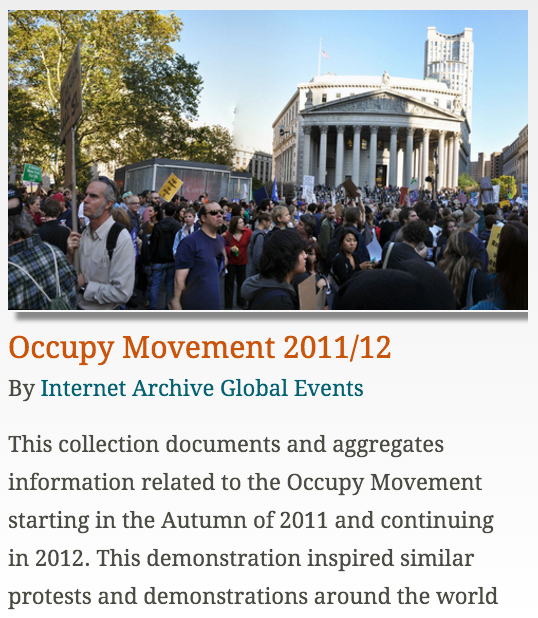} 
\label{occupy}
\end{wrapfigure}

The collection \textit{Occupy Movement 2011/12} by \textit{Internet
  Archive Global Events} started in November 2011 with members from
the Archive-It team and a large web archiving community
collaboratively identifying and capturing websites related to the
\textit{``Occupy Movement''}. The movement started in Autumn 
2011 and continued in 2012. The collection includes movement-wide or
city specific occupy websites, blogs, social media sites and news
articles from alternating or traditional media outlets.

The selection of seeds involved public participation by content
curators and individuals, as well as crawling of websites from
community generated feeds (commencing on December 2011) with
\textit{933 seed URLs} selected (e.g. ``Occupy Feeds''). In the
subsequent years the collection was crawled frequently, capturing
about \textit{26 million documents} and archiving over \textit{900GB}
of data.

The Archive-It team is also collecting web content on global events at
risk of disappearing such as \textit{``Earthquake in Haiti'', ``North
  Africa \& Middle East 2011-2013'', ``Ukraine Conflict'', ``2013
  Boston Marathon Bombing''} and others.

\endgroup

The collection consists of seeds organized into groups such as blogs,
international news sites and articles, other sites and social
media. An interesting
analysis\footnote{\url{https://archive-it.org/blog/post/only-41-of-occupy-movement-urls-accessible-on-live-web/}}
of the collection was carried out in 2014 to see how much of the seed
contents still live on the web. The seeds were categorized into news
articles, social media URLs, and ``movement sites''. Only 41\% of the
582 ``movement sites'', 85\% of 203 social media URLs and 90\% of 163
news articles are still alive on the web. All ``movement sites'' that
were no longer available on the web, were either showing 404 errors or
taken down by cyber squatters.

\section{ArchiveWeb: Functionalities and Evaluation}\label{sec:System}

ArchiveWeb facilitates collaborative exploration of multiple focused
web archive collections. The features provided by the system to
support this exploration are: \textit{searching} across multiple
collections in conjunction with the live web, \textit{grouping} of
resources for creating, merging or expanding collections from existing
ones and  \textit{enrichment} of resources in existing
collections using comments and tags. 

In order to collect relevant feedback, we carried out a task-based
evaluation of ArchiveWeb involving experts from different university libraries and archiving
institutions, including a follow-up qualitative interview. 

For the evaluation, we imported some publicly
accessible web archive collections from Archive-It into ArchiveWeb in order to
test the potential of the system to support collaborative work
with such collections. We invited eight experts from university
libraries and archiving organizations
to evaluate the system. Evaluators were from the following
institutions: Columbia University, University of Toronto, Cornell
University, Stanford Libraries, National Museum of Women Arts, New
York Arts Resources Consortium and the Internet Archive. We asked them to
carry out two sets of tasks both \textit{individual} and
\textit{collaborative}. The \textit{individual tasks} included
creating a sub-collection from an existing collection of their
institution, enriching an existing collection with additional relevant
resources from the live web and annotating resources with information
about why they should be included and how often they should be
archived. The \textit{collaborative tasks} involved selecting ten
featured resources from their collections, including them into a joint
group shared with all other evaluators, and discussing the reasons why such resources were
included. The final task involved creating a new collection about a
shared topic among the evaluators, searching for relevant seed URLs
for this collection using ArchiveWeb, and agreeing with the rest of the
evaluators about which seeds should be included, what should be the
archiving frequency and why.

The evaluation period spanned about three weeks, after which we
invited the evaluators to give us feedback through a questionnaire and
through follow-up interviews to fully understand their experience.  

The following sections give an
overview over the main insights based on both quantitative analysis of
query logs and qualitative feedback from the interviews, along with
the description of the system functionalities.

\subsection{Searching for Resources in Collections and on the Web}\label{sec:search}
\begin{figure}[hbp]
\centering
\includegraphics[width=0.95\textwidth]{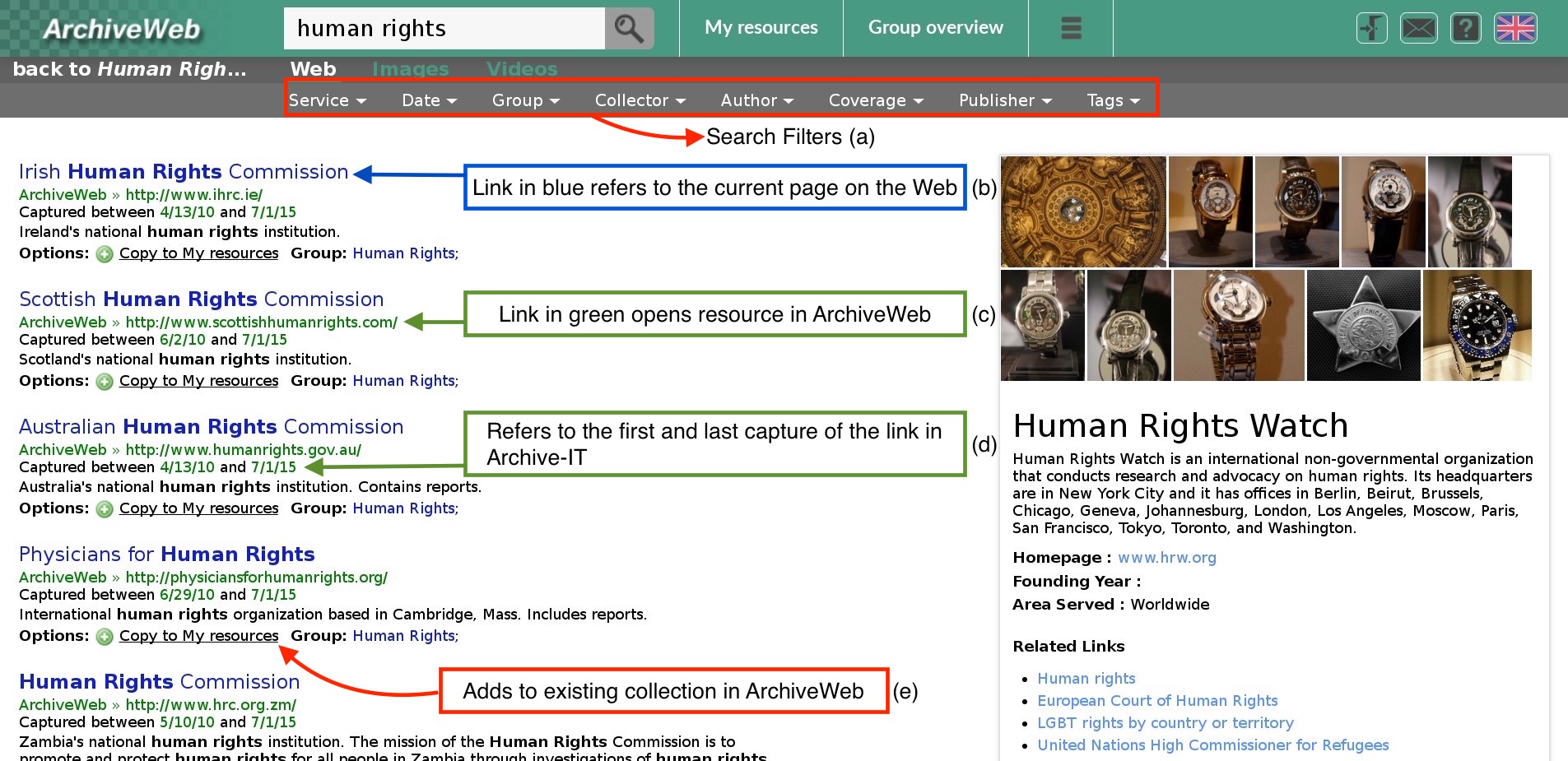}
\caption{Search Page}
\label{search}
\end{figure}
ArchiveWeb provides a keyword based search system that
returns results from Archive-It collections as well as from the live
web (using the Bing Search
API\footnote{\url{http://datamarket.azure.com/dataset/bing/search}}). If
users search for a keyword (e.g. ``human rights''), ArchiveWeb returns
a list of results from these sources, indicating whether the resource
comes from a specific Archive-It collection or from the live web
(Fig. 2b). The results being returned from Archive-It collections are
based on a full metadata index for these collections (a fulltext index
of all content is planned for the next version). Besides webpages,
also images (from Bing, Flickr, Ipernity) and videos (from YouTube,
TED, Yovisto, Vimeo) can be searched, but not yet from Archive-It
collections.  For Archive-It collections each search result provides a
pointer to the resource item saved in ArchiveWeb (Fig. 2c) and
displays information about when it was captured in
Archive-It(Fig. 2d).  Live web results are not yet saved in ArchiveWeb
and can be copied by the user into a group (Fig. 2e); in this case,
the Option window provides details about when/if this page has been
indexed by the Internet Archive, using the Wayback CDX server
API\footnote{\url{https://github.com/internetarchive/wayback/tree/master/wayback-cdx-server}}.

\begin{figure}[!ht]
\includegraphics[width=\textwidth]{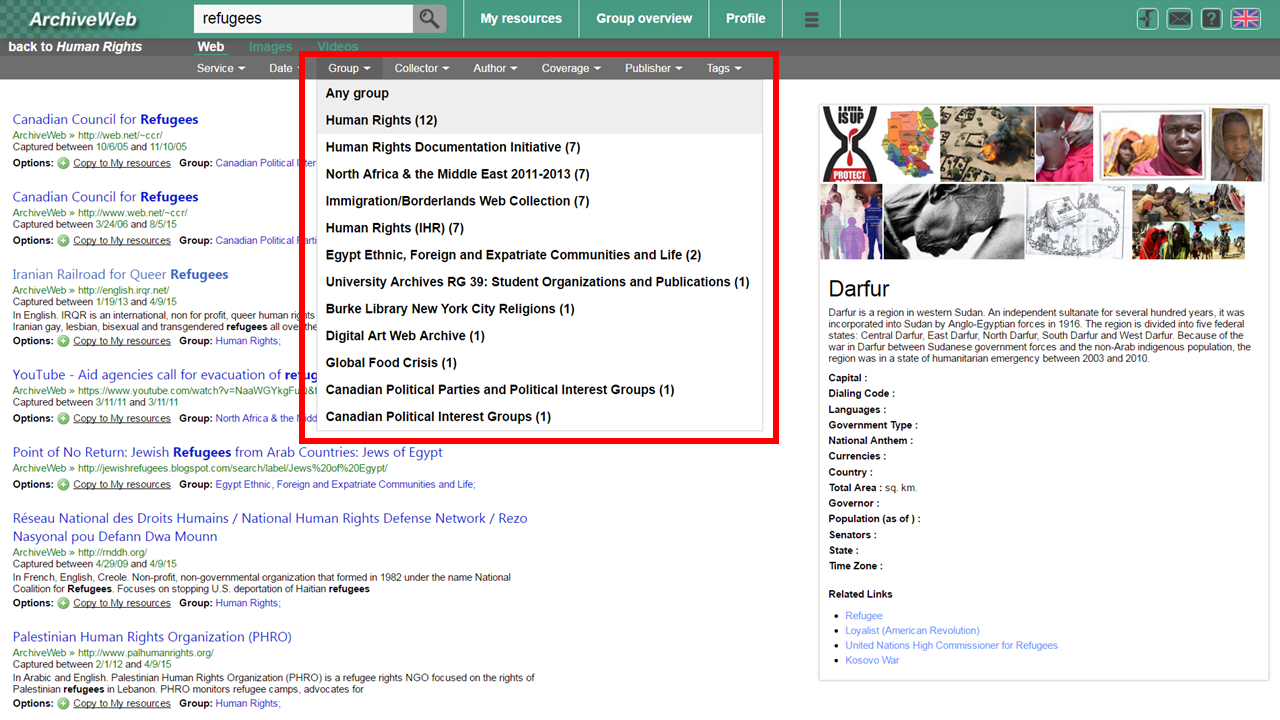}
\caption{Refinement through Faceted Search Filters}
\label{facet}
\end{figure}

Search results can be refined using faceted search
filters visible below the search box (Fig. \ref{facet}). For example,
for our evaluation it was important to provide filters to show only
the resources in a specific collection (Filter: \textit{Group}) or
that were archived by a specific institution (Filter:
\textit{Collector}).

\textbf{Feedback} The integration of searching across web archive
collections as well as the live web provides the ability to suggest additional
material that could be archived. This is a new feature not
present in existing systems used for curation/creation of web
archive collections, and it was much appreciated by the evaluators. 
The search filters
mainly used while searching were ``Author'', ``Tags'', ``Group'' and
``Service''. These filters were mainly used to understand how the
results would change with different important filter options while
either searching across Archive-It collections or the live web.

Suggestions we received included: (i) providing a transparent
explanation of how the relevance ranking is determined, (ii) providing
the total number of captures of a page along with the capturing period
in order to determine the popularity of an archived result, (iii)
providing as default just search results from Archive-It collections
with the option to expand the results to the live web in order to
facilitate curators who expect to see only archived resources in the first step.

\subsection{Organizing and Extending Resource Collections}\label{sec:group}

ArchiveWeb provides the functionality to organize resources into
collections (groups of resources) according to clearly defined and coherent
themes/topics. This functionality allows working with existing groups, creating new
collections/groups and sub-groups, adding new resources to groups, and 
moving resources between groups.

\begin{figure}[htbp]
\includegraphics[width=\textwidth]{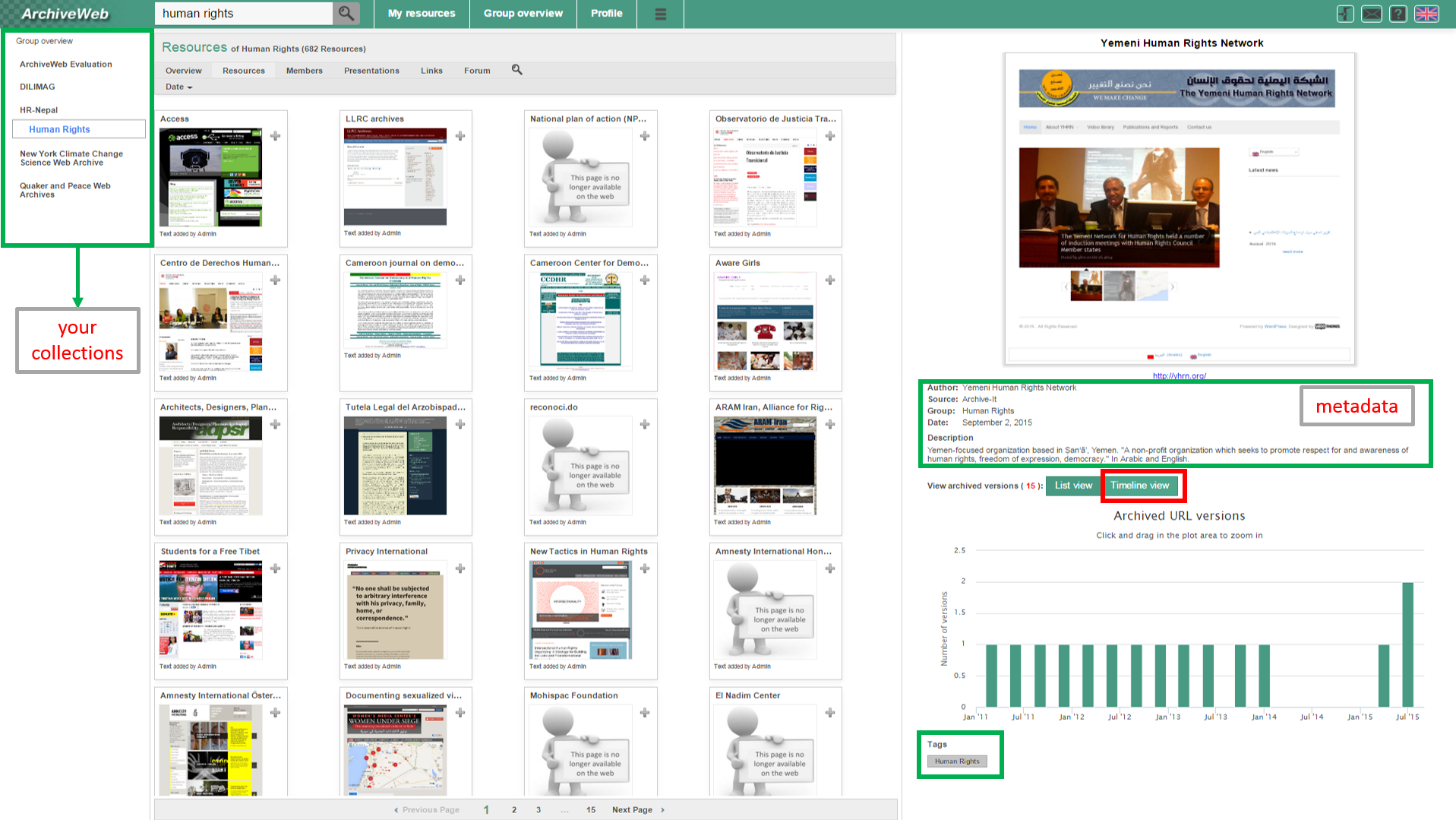}
\caption{Group Resources Interface}
\label{group}
\end{figure}

A group overview interface allows browsing through existing collections
available within ArchiveWeb and the collections that a user has
created or joined. Descriptions for every collection provide
information about the topic/theme and what kind of resources it
contains. For Archive-It collections, the
description reports the details available on the public Archive-It
interface along with a reference link to the specific collection in the
ArchiveWeb platform.

After joining specific groups/collections users have the opportunity
to edit metadata of existing resources and to contribute new resources
from other existing collections or from the live web. The ArchiveWeb
collections, derived directly from their Archive-It counterparts, are
``read-only''.  Users can browse through the resources of each
collection using the advanced visualization and exploration
functionalities of ArchiveWeb, but they cannot change the original
collection. They can create a copy of the entire ``read-only'' group,
or add individual resources to their own collections by selecting
resources individually, as well as merge multiple existing collections
into one new collection.

Users can also organize their resources into sub-groups within
collections in order to group resources that are related to a similar
subtopic. Resources can be uploaded to a collection either from the
desktop or by suggesting a URL. Each group has a specific interface
that visualizes all thumbnails of the most recent snapshot of each
resource when it was added to the group (Fig. \ref{group}). Resources
that no longer exist on the web, or do not have a redirect available,
display a thumbnail with the message ``The page is no longer available
on the web'' (we are currently implementing the functionality to
upload an earlier thumbnail of the page taken from the Internet
Archive Wayback Machine). The overview interface within a group
provides a summary of the activities of various members of the group,
including the actions of new added resources, resources which were
edited or deleted, and users who have joined/left the group.

\textbf{Feedback} One of the evaluators stated, that
\textit{``ArchiveWeb increases curatorial functionality over that of
  Archive-It, e.g.: on-the-fly creation of groups, moving resources
  between groups and easily annotating resources''}. Several of the
new functionalities of ArchiveWeb were positively highlighted by the
evaluators including (i) the ability to curate new arbitrary collections
of seed records from across multiple web archive collections, (ii) the
possibility of having resources exist at multiple levels (e.g. personal, group,
subfolder), and (iii) the ability to create collaborative
collections with colleagues from various institutions.

\subsection{Enriching Resource Metadata and Discussing Resources}\label{sec:enriching}

After a resource is added to an ArchiveWeb collection the resource
can be enriched with additional comments and tags (in addition to the
metadata already provided). The comments on a resource can be used to
discuss as to why this resource has been chosen as a seed for a
collection, and to decide upon the crawl frequency and crawl depth. By
exchanging comments, collaborators can discuss the relevance of a
suggested seed resource for a particular collection. The use of tags
helps categorize or label a resource with subjects or topics
covered by the resource, making it possible to browse collections by
filtering based on certain tags. Users can also edit metadata such as title, description and author fields.

The system allows users to archive a single page or a website
(resource) by clicking on the ``\textit{Archive Now}'' button which
sends a request to the Wayback Machine to archive it. This
functionality is similar to the ``\textit{Save Page Now}'' feature of
the Wayback machine, but it facilitates users to gather captures
easily as they work within the system. All the captures from
Archive-It as well as the wayback captures added using the
``\textit{Archive Now}'' functionality are visualized both as a
\textit{list} and in a \textit{timeline} view to help users navigate
through the different archived versions that are available. Figure
\ref{group} shows a timeline view on the bottom right.

\textbf{Feedback} One of the new functionalities that was positively
highlighted by the evaluators is the possibility to add
schema-agnostic classification and other information using tags.
Other appreciated functionalities were (i) the ability to use notes/comments in order to highlight the motivations for collecting and
to discuss the capture frequency, highly useful when
developing collaborative collections across institutions, (ii) \textit{list} and \textit{timeline} visualizations were
considered useful to explore individual seed records.

Metadata fields which were generally modified while editing a resource included
\textit{title} and \textit{description} (as not all of them are 
automatically filled appropriately while uploading a URL to a group),
and the \textit{author} field which is often missing from the original
metadata. An interesting suggestion from one of the evaluators was to add open/customizable
metadata fields besides \textit{title}, \textit{description} and
\textit{author} as they would be needed if curators from different
institutions build collections together. Tagging is helpful but not
considered a substitute for using custom field names and corresponding
values.

\subsection{Future Improvements}

The evaluators also provided additional suggestions for
functionalities that are missing in the ArchiveWeb system which
we plan to incorporate in the next release, for example:
\begin{description}
\item[Exporting Resources] - the ability to export
  resources and metadata from collections for research or
  usage outside of ArchiveWeb.
\item[Bulk Operations] - the ability to select multiple resources by clicking, searching or
  filtering, and then adding them to a collection. The system should also support bulk editing of resources within a collection
  such as collective tagging (e.g. to assign a tag to multiple seed records),
  as well as tagging similar resources across multiple collections and
  then allowing the option to group them together into a new
  collection.
\item[Advanced Search] - the ability to limit the search within
  certain collections, specific domains or paths, and to search within
  the title only. We also plan to incorporate fulltext search of
  Archive-It websites instead of just the metadata, even though not
  all evaluators realized that this functionality was not
  provided, as rich metadata descriptions are available for many
  collections.
\end{description}

\section{Conclusion}\label{sec:Conclusion}

In this paper we discussed the ArchiveWeb system which supports
collaborative exploration of web archive collections. We provided a
description of the main features of the system such as (i) searching
across multiple collections as well as the live web, (ii) grouping of
resources for creating new collections or merging existing ones, as
well as (iii) collaborative enrichment of resources using comments and
tags.

The system has been developed based on an iterative evaluation-driven
design based research approach. Starting from a platform which already
supported collaborative search and sharing of web resources,
ArchiveWeb was designed to address expert users' requirements
(e.g. librarians and curators in archiving institutions). The
resulting ArchiveWeb prototype, fully functional, was evaluated
through a task-based evaluation study carried out with the same
experts who participated in the preliminary investigation. After the
quantitative analysis of the logs and the qualitative feedback from
the evaluations, we are now incorporating new features such as
exporting resources, bulk editing operations and advanced search which
will be available in the next release of the system.

\section*{Acknowledgments}\label{sec:Acknowledgments}

We thank Jefferson Bailey from the Internet Archive who provided us
with the contacts to his colleagues at university libraries and
archiving institutions. We are also grateful to all experts, who
participated with enthusiasm in our evaluation, providing valuable
feedback and useful suggestions to improve the system. This
work was partially funded by the European commission in the context of
the ALEXANDRIA project (ERC advanced grant no 339233).

\bibliographystyle{splncs03}

\bibliography{ArchiveWeb}

\end{document}